\documentclass[conference]{IEEEtran}
\IEEEoverridecommandlockouts
\usepackage{cite}
\usepackage{amsmath,amssymb,amsfonts}
\usepackage{algorithmic}
\usepackage{graphicx}
\usepackage{textcomp}
\usepackage{xcolor}
\usepackage{subfigure}
\usepackage{algorithm2e}
\usepackage{array}
\usepackage{tabularx}
\usepackage{adjustbox}
\usepackage{multirow}
\usepackage{nicematrix}
\usepackage{makecell}

\newcommand{\bda}[1]{$B_{D/A}$#1}

\def\BibTeX{{\rm B\kern-.05em{\sc i\kern-.025em b}\kern-.08em
    T\kern-.1667em\lower.7ex\hbox{E}\kern-.125emX}}

\begin{document}

\bstctlcite{IEEEexample:BSTcontrol}

\title{OSA-HCIM: \underline{O}n-The-Fly \underline{S}aliency-\underline{A}ware \underline{H}ybrid SRAM \underline{CIM} with Dynamic Precision Configuration\\}


\author{\IEEEauthorblockN{Yung-Chin Chen\IEEEauthorrefmark{1}\IEEEauthorrefmark{2}, 
Shimpei Ando\IEEEauthorrefmark{1}, 
Daichi Fujiki\IEEEauthorrefmark{1}, 
Shinya Takamaeda-Yamazaki\IEEEauthorrefmark{3},
Kentaro Yoshioka\IEEEauthorrefmark{1}}
    \IEEEauthorblockA{
    \IEEEauthorrefmark{1}Keio University, Japan,
    \IEEEauthorrefmark{2}National Taiwan University, Taiwan,
    \IEEEauthorrefmark{3}The University of Tokyo, Japan
    }
    jim.chen.work@gmail.com, \{shimpeiando, dfujiki\}@keio.jp, shinya@is.s.u-tokyo.ac.jp, kyoshioka47@keio.jp
    }

\maketitle

\begin{abstract}
Computing-in-Memory (CIM) has shown great potential for enhancing efficiency and performance for deep neural networks (DNNs).
However, the lack of flexibility in CIM leads to an unnecessary expenditure of computational resources on less critical operations, and a diminished Signal-to-Noise Ratio (SNR) when handling more complex tasks, significantly hindering the overall performance. Hence,  we focus on the integration of CIM with Saliency-Aware Computing---a paradigm that dynamically tailors computing precision based on the importance of each input.
We propose On-the-fly Saliency-Aware Hybrid CIM (OSA-HCIM) offering three primary contributions: (1) On-the-fly Saliency-Aware (OSA) precision configuration scheme, which dynamically sets the precision of each multiply-and-accumulate (MAC) operation based on its saliency, (2) Hybrid CIM Array (HCIMA), which enables simultaneous operation of digital-domain CIM (DCIM) and analog-domain CIM (ACIM) via split-port 6T SRAM,
and (3) an integrated framework combining OSA and HCIMA to fulfill diverse accuracy and power demands.

Implemented on a 65nm CMOS process, OSA-HCIM demonstrates an exceptional balance between accuracy and resource utilization. Notably, it is the first CIM design to incorporate a dynamic digital-to-analog boundary, providing unprecedented flexibility for saliency-aware computing. OSA-HCIM achieves a 1.95x enhancement in energy efficiency, while maintaining minimal accuracy loss compared to DCIM when tested on CIFAR100 dataset.
\end{abstract}


\section{Introduction}
The massive data communication between memory and computing units presents a significant challenge to the efficient hardware acceleration of Deep Neural Networks (DNNs).
This often results in increased energy consumption and protracted processing times. A promising solution to this issue is Computing-in-Memory (CIM), a technology that integrates computational functions into the memory array, thereby reducing data movement and significantly improving energy efficiency. 
Though general-purpose CIM technology has matured with extensive research and macro-level implementations \cite{yue202115}, there is an ongoing shift towards developing CIM solutions that exploit network characteristics to achieve further performance breakthroughs.
Such endeavors include 
leveraging DNN data distributions \cite{yue202328nm} and designing for specific network architectures \cite{wang202328nm}.

\begin{figure}[tbp]
\centerline{\includegraphics[width=9cm]{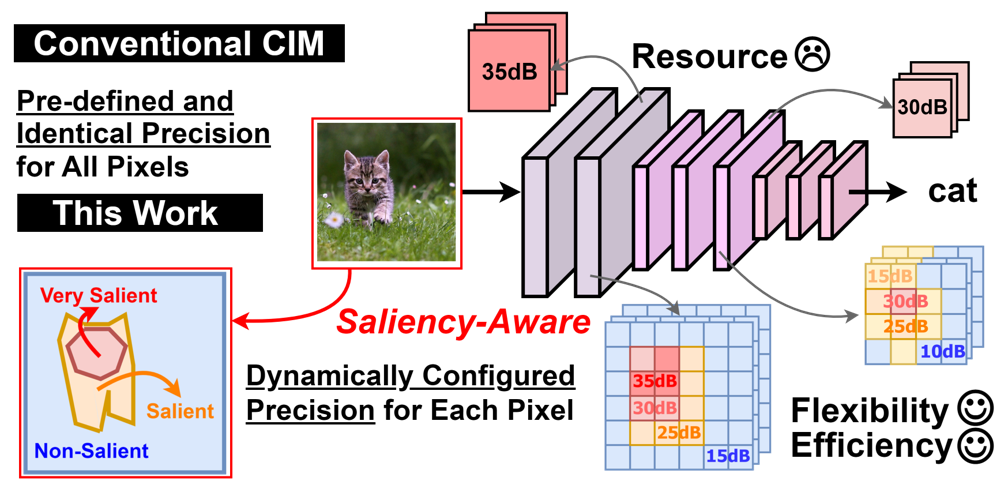}}
\caption{Motivation of Saliency-Aware Computing. 
Each input pixel has a distinct impact on the output result of the DNN, necessitating different precision configurations. By dynamically employing high signal-to-noise ratio (SNR) precision for salient inputs and low SNR precision for non-salient inputs, efficiency can be enhanced.
}
\label{saliency}
\vspace{-4mm}
\end{figure}

This paper explores \textit{Saliency-Aware Computing}, a software-driven mixed-precision computation paradigm which dynamically tailors computing precision based on the saliency, or the importance, of each input pixel. 
Figure \ref{saliency} shows the application of Saliency-Aware Computing to an image recognition task, using an image of a cat as an example. It is crucial to note that not all pixels hold equal importance; for instance, the pixels composing the cat's face and body are vital for recognition, while the background pixels depicting grass and flowers are largely irrelevant. By \textit{dynamically} adjusting the computation precision in line with input saliency, computational resources can be effectively utilized. Specifically, we focus on high-precision computation for salient inputs that greatly impact classification accuracy, and lower-precision computation for non-salient inputs that contribute less. This selective allocation of computational resources allows for significant reductions in computational costs while maintaining high accuracy levels.


Despite the potential of saliency-aware computing to significantly enhance CIM performance, conventional CIMs lack the functionality to enable dynamic computing precision configuration.
One contributing factor is the limitations within the vector-accumulation circuitry. For example, Digital-domain CIM (DCIM) necessitates excessive computation resources \cite{chih202116, TSMC2022}, and Analog-domain CIM (ACIM) often compromises accuracy \cite{lee2021chargesharing8T, lee2022lowcost}. 
Recent studies have attempted to integrate the digital and analog circuitries within the same macro unit \cite{chen2022charge, wu202322nm}. However, these hybrid schemes rely on fixed digital-to-analog configurations, which poses limitations on flexibility. To fully leverage the benefits of saliency-aware computing, a CIM design should have the capacity to 1) configure its multiply-and-accumulate (MAC) precision real-time with the saliency levels of input activation, and 2) allow dynamic configuration of the digital-to-analog computation ratios.
To tackle these challenges, we introduce OSA-HCIM: a software-hardware co-designed hybrid CIM architecture. Differing from previous works, our approach synergistically combines the merits of software and hardware strategies. We specifically leverage the largely untapped potential of saliency-aware computing in the realm of CIM, and address the current limitations in dynamically configuring precision in hybrid CIMs.
OSA-HCIM is characterized by three key features:
(1) \textbf{Software Realm}: We introduce an On-the-fly Saliency-Aware (OSA) precision configuration scheme, providing numerous precision configurations for the MAC operation based on the online-evaluated saliency.
(2) \textbf{Hardware Realm}: We propose a Hybrid CIM Array (HCIMA) capable of performing both bit-serial DCIM and bit-parallel ACIM concurrently using split-port 6T SRAM.
(3) \textbf{Software-Hardware Co-design}: We present a comprehensive framework that integrates OSA scheme into HCIMA using a near-memory On-the-fly Saliency Evaluator (OSE) to effectively address various accuracy and efficiency requirements. 
The 64 $\times$ 144 6T SRAM OSA-HCIM macro, implemented using 65 nm CMOS technology, stands out as the first CIM work to harness input saliency and incorporate a dynamic digital-to-analog computing boundary. The result demonstrates a 1.95x enhancement in energy efficiency while preserving similar levels of accuracy when compared to the full-digital approach on the CIFAR100 dataset.



\section{Preliminaries}

\subsection{Saliency-Aware Compute}
The concept of saliency, which represents the varying importance of inputs, has been extensively utilized in various computer vision tasks such as video compression \cite{hadizadeh2013saliency}, super resolution \cite{sadaka2011efficient}, and object recognition \cite{ren2013region}.
This feature can also be leveraged in DNN tasks, as the contribution of each input activation to the final output result can vary significantly.
For instance in Fig. \ref{saliency}, pixels containing the object of interest are more salient than the background pixels. To address this, saliency-aware computation have been developed. The computational precision is adjusted based on the input's saliency, where higher precision to salient pixels and lower precision to less significant ones is assigned. This strategy can reduce computational costs while minimally impacting the accuracy results.

Precision Gating (PG) \cite{PG} is a dual-precision software technique that assesses saliency by performing MAC operations between high-order input bits and weights, while DRQ \cite{DRQ} is a hardware architecture that calculates saliency by employing a mean filter on the input. However, these approaches have limitations: (1) support only dual precision configurations, providing limited tradeoff efficacy, and (2) lack memory-centric hardware solution.

\subsection{Digital, Analog, and Hybrid CIM}


CIM architectures incorporate computing circuits within memory cells to perform MAC operations. Digital-domain CIM (DCIM) utilizes bulky digital adder trees (DATs) to perform loss-free accumulations \cite{chih202116, TSMC2022}, while analog-domain CIM (ACIM) employs analog accumulation technique, such as charge-sharing, along with an ADC conversion to achieve efficient but less accurate accumulation \cite{lee2021chargesharing8T, lee2022lowcost}.
To balance accuracy and resource consumption, recent research has explored the use of hybrid accumulation strategies that adopt both digital and analog schemes. This approach bifurcates the multi-bit MAC operation into DCIM and ACIM based on the input or output order using a pre-set threshold. Existing works either partition MAC operations into separated analog and digital CIM blocks based on pre-defined constraints \cite{rashed2021hybrid}, or implement both DCIM and ACIM within a macro with hard-wired boundary between digital and analog MAC computation \cite{chen2022charge, wu202322nm}.

Nonetheless, the boundaries between digital and analog computations in these works are predetermined, and therefore lack adaptability to the varying precision requirements of each input. As a result, accuracy suffers due to analog noise when the input is more salient, while excessive resources are allocated for trivial inputs, degrading efficiency. To overcome this challenge, there is a need for a dynamically-configurable hybrid CIM capable of adjusting the digital and analog ratio according to the nature of the input patterns.



\section{Software Realm: On-the-fly Saliency-Aware (OSA) precision configuration scheme}
We propose the On-the-fly Saliency-Aware (OSA) precision configuration scheme, which introduces a dynamic computing precision mechanism optimized for CIM architectures. By evaluating the saliency of each input in real-time, OSA dynamically adjusts the ratio of digital and analog computation, thus enabling efficient resource utilization.

The OSA scheme operates within a standard CIM system that is optimized for binary (or 1-bit) MAC operations. To facilitate multi-bit MAC operations in such CIMs, $a$-bit input activations $\vec{A}[a-1: 0]$ and $w$-bit weights $\vec{W}[w-1: 0]$ are decomposed into $w \times a$ 1-bit MACs with output order $k=i+j$ between 1-bit weights $\vec{W}[i]$ and 1-bit activation $\vec{A}[j]$, where $i$ ($0\le i \le w-1$) and $j$ ($0\le j \le a-1$) indicate the bit order of weight and activation, respectively \cite{cyyao}. 
This decomposition is described in the equation below: 
\begin{equation}
    \mathit{MAC}(\vec{A}, \vec{W})=\sum_{i=0}^{a-1}\sum_{j=0}^{w-1} 2^{i+j} \cdot \mathit{MAC}(\vec{A[i]}, \vec{W[j]}) 
\end{equation}

\begin{figure*}[tbp]%
\centering
\subfigure[]{%
\label{sw_framework}%
\includegraphics[height=5cm]{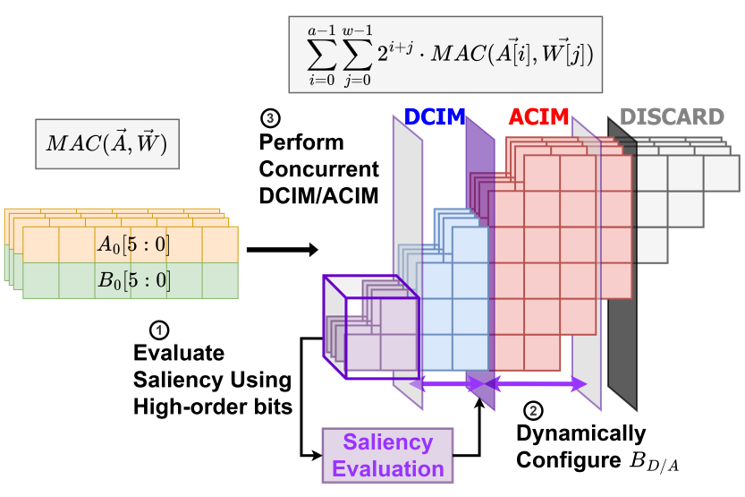}}%
\qquad
\subfigure[]{%
\label{overview}%
\includegraphics[height=6cm]{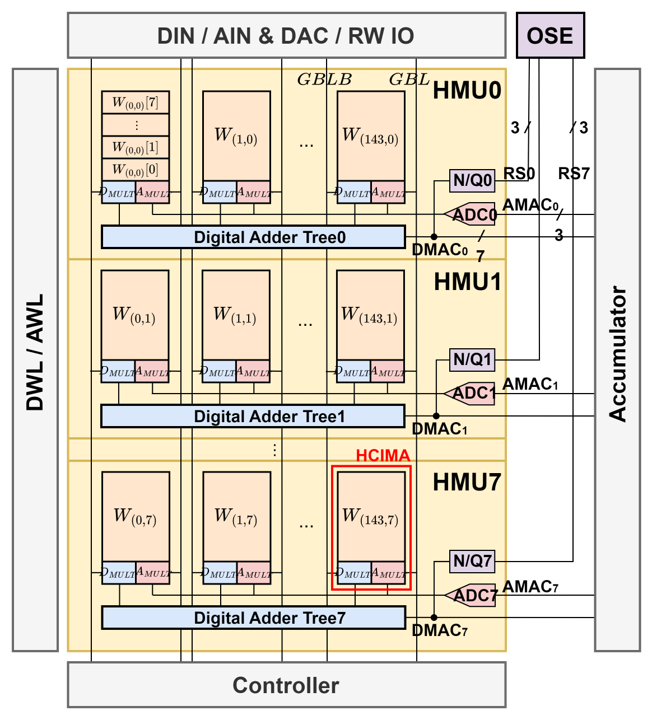}}%
\qquad
\subfigure[]{%
\label{HCIMA}%
\includegraphics[height=6cm]{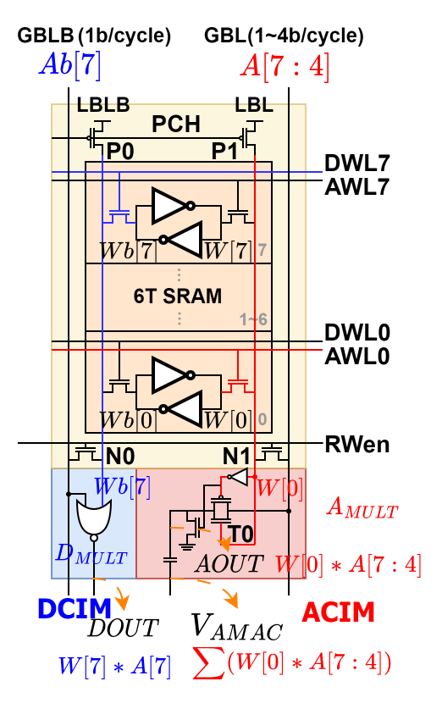}}%

\caption{(a) On-the-fly Saliency-Aware (OSA) Precision Configuration Scheme. \bda{} is dynamically configurable, allowing for a tradeoff between accuracy and efficiency based on input saliency. (b) The overall OSA-HCIM architecture. (c) Schematic of HCIMA. 
}
\vspace{-4mm}
\end{figure*}

Figure \ref{sw_framework} illustrates the proposed OSA precision configuration scheme integrated into the decomposed multi-bit Hybrid CIM MAC operation. As an example, we consider a case where $w=a=6$, with 36 distinct 1-bit MACs in total. To evaluate the saliency of the entire multi-bit MAC operation, the first $s$ ($s$=2 in Fig. \ref{sw_framework}) highest-order 1-bit MACs, corresponding to MACs with output order $k=w+a-2 \sim w+a-1-s$, are computed precisely using DCIM. Subsequently, the output saliency of the multi-bit MAC is estimated based on these high-order 1-bit MAC results, which in turn determines the digital-to-analog boundary, \bda. Consequently, each 1-bit MAC is categorized into \textit{digital mode}, \textit{analog mode}, or \textit{discard} based on its output order $k$. 
MACs with $k \ge B_{D/A}$ are computed with no accuracy compromise using DCIM, while those with $B_{D/A}-4 \le k < B_{D/A}$ are computed with energy-efficient ACIM, albeit with noise incorporation. To further save power, MACs with $k < B_{D/A}-4$ are discarded due to  their negligible impact on the final output. By dynamically configuring \bda{}, a unique precision configuration can be achieved, which enables the exploration of the optimal accuracy-efficiency tradeoff.

\section{Hardware Realm: OSA-HCIM Architecture}

\subsection{OSA-HCIM Architecture Overview}

The proposed OSA-HCIM hardware architecture features concurrent CIM operation across both the digital and analog domains. OSA-HCIM supports parallel execution within these hybrid domains, while providing the requisite flexibility for implementing our OSA precision configuration scheme. The culmination of these features leads to a substantial enhancement in computational performance and efficiency.

Figure \ref{overview} provides an architecture overview of OSA-HCIM. It is composed of the OSA-HCIM macro and its peripherals, including an On-the-fly Saliency Evaluator (OSE), an Accumulator, Digital/Analog WL Drivers (DWL/AWL), Digital Input Drivers (DIN), Analog Input Drivers (AIN) along with Digital-to-Analog Converters (DACs), a read/write IO (R/W IO), and a Controller. The 64b $\times$ 144b OSA-HCIM macro contains 8 Hybrid MAC Units (HMU), each of which comprises 144 Hybrid CIM Arrays (HCIMA), a Digital Adder Tree (DAT), a Normalization-and-Quantization Unit (N/Q), and a 3-bit SAR-ADC.

To perform \textit{CIM} operations, OSA-HCIM executes both DCIM and ACIM within the HCIMA simultaneously.  The MAC results of DCIM and ACIM, referred to as DMAC and AMAC, are then combined by shifting and adding them in the accumulator, resulting in the final multi-bit MAC output. Further details regarding the functioning of DCIM and ACIM are provided below.

\noindent \textbf{Digital CIM (DCIM):} 
The digital input activations are dispatched to GBLBs in a bit-serial manner through DINs. Within each HCIMA, the digital circuitry carries out a bitwise multiplication of the stored weight and input, generating the Digital Output (DOUT). The DOUTs from the 144 columns are then aggregated by the Digital Adder Tree (DAT), yielding a 7-bit output DMAC.

\noindent \textbf{Analog CIM (ACIM):} 
Analog input activations of 1 to 4 bits are initially converted into analog voltages on GBLs utilizing DACs, and subsequently driven through AINs. The DAC is implemented via a switch matrix between reference voltages, enabling flexibility in bit-precision and supporting adaptable mapping of the analog MAC operations. In every HCIMA, the analog circuitry undertakes the multiplication of bit-serial weight and bit-parallel activation to produce AOUT. These AOUTs are summed using charge-sharing and converted to a 3-bit output AMAC via the SAR-ADC. Here, extremely low ADC precision is employed since ACIM is exclusively used for computing less significant data.

When carrying out the multi-bit MAC operations, the OSA-HCIM macro switches between two modes, namely, the Saliency Evaluation Mode and the Computing Mode. 

\noindent \textbf{Saliency Evaluation Mode:} 
The purpose of this mode is to evaluate the saliency of the entire MAC operation utilizing few highest-order 1-bit MACs (Step 1 of Fig. \ref{sw_framework}). In this mode, the DMACs are quantized to 3-bit via the Normalization-and-Quantization Unit (N/Q) and sent to the On-the-fly Saliency Evaluator (OSE). The OSE evaluates the saliency based on these quantized DMACs (Step 2 of Fig. \ref{sw_framework}).

\noindent \textbf{Computing Mode:} 
Once the saliency is evaluated, the OSA-HCIM macro transitions to the Computing Mode. The role of this mode is to execute the remaining MAC operations based on the saliency score obtained from the Saliency Evaluation Mode (Step 3 of Fig. \ref{sw_framework}). Since the OSE sets the digital-to-analog boundary to optimum \bda{}, the accuracy and resource consumption will be balanced.

\begin{figure}[tbp]%
\centering
\subfigure[]{%
\label{OSE}%
\includegraphics[width=4.1cm]{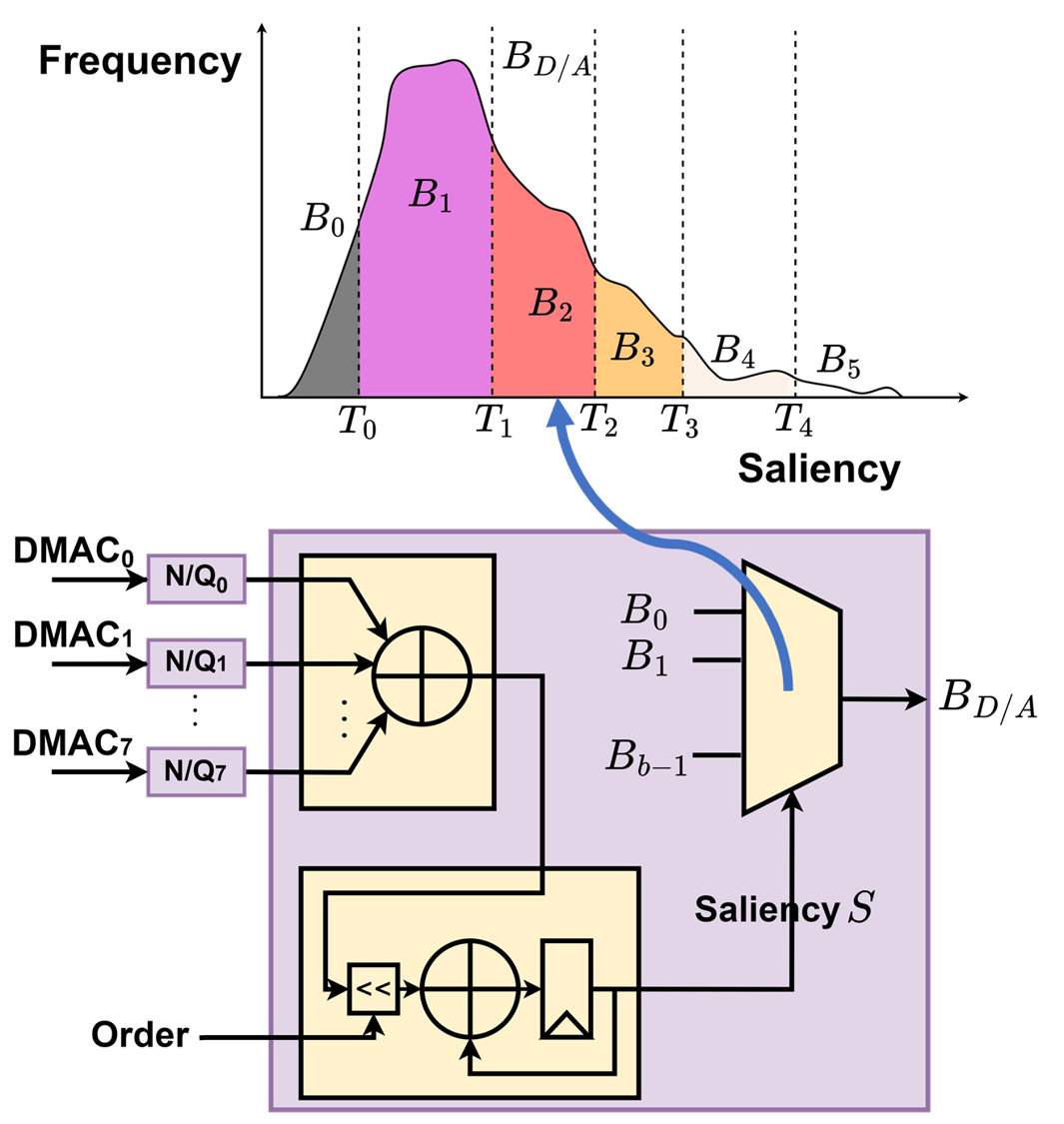}}
\subfigure[]{%
\label{thres algo}%
\includegraphics[width=2.8cm]{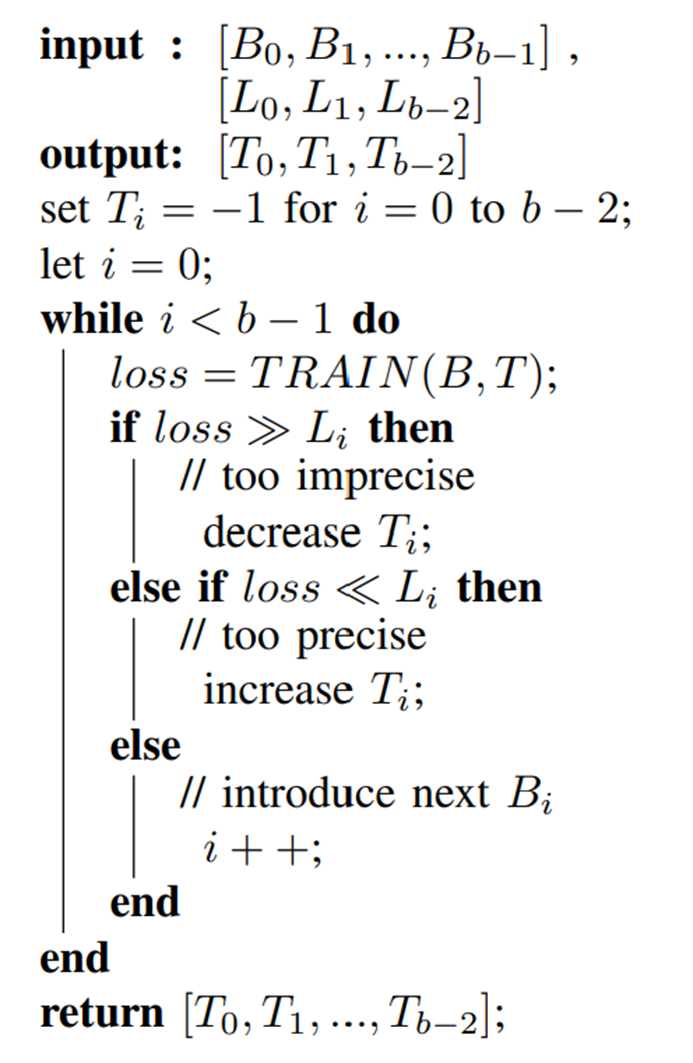}}%
\caption{The architecture of On-the-fly Saliency Evaluator (OSE). (b) The algorithm for determining \bda{} thresholds $T$ for OSE. $T$ can be pre-trained without degrading inference performance.
}
\vspace{-4mm}
\end{figure}

\subsection{Hybrid CIM Array (HCIMA)}

Figure \ref{HCIMA} illustrates the structure of our proposed Hybrid CIM Array (HCIMA), which is a novel CIM structure capable of \textit{simultaneously} executing DCIM and ACIM, thereby effectively doubling the CIM throughput. 
The structure of HCIMA is composed of eight split-port 6T SRAMs, a pair of NMOS transistors (N0, N1) for conventional read/write operations during \textit{RW} state, two PMOS transistors (P0, P1) responsible for the precharge of LBL and LBLB, and both digital ($D_{MULT}$) and analog ($A_{MULT}$) multipliers. Each HCIMA allows for the storage of either one 8-bit weight or two 4-bit weights.
In \textit{RW} state, the $RWen$ signal is elevated to activate N0 and N1. Both DWL and AWL from the target row are engaged for the standard SRAM read/write operations. Conversely, for \textit{CIM} operation, the $PCH$ signal is initially set down to enable P0/P1 to precharge LBLB/LBL. 

Owing to the split-port readout scheme, different weights can be independently read on LBL and LBLB, thereby facilitating simultaneous digital and analog computations. Concurrently, GBLB and GBL transmit 1-bit inverted digital activation and 1$\sim$4-bit analog activation to $D_{MULT}$ and $A_{MULT}$ respectively, instigating the multiplication of weight and input activation.
For example, in the case shown in Fig. \ref{HCIMA}, DWL7 and AWL0 are activated to readout $Wb[7]$ on LBLB and $W[0]$ on LBL. Meanwhile, $Ab[7]$ and $A[7:4]$ are sent via GBLB and GBL respectively. Subsequently, DOUT outputs the result of $W[7] \times A[7]$, and AOUT outputs the result of $W[0] \times A[7:4]$. 
The digital bit-serial multiplication of $D_{MULT}$ is realized with a simple NOR gate, while the analog bit-parallel multiplication of $A_{MULT}$ is realized with a transmission gate (T0), alongside an NMOS transistor (N2) for pull-down. 
\section{Software-Hardware Co-design: OSA-HCIM Framework}

The OSA-HCIM framework integrates the OSA precision configuration scheme of the software realm into the OSA-HCIM macro through the use of near-memory On-the-fly Saliency Evaluator (OSE). 
This framework also performs workload allocation to instigate DCIM and ACIM operations effectively in HCIMA. The details of OSE and workload allocation method are discussed below.

\begin{figure}[tbp]%
\centering
\subfigure[]{%
\label{workload_allocate}%
\includegraphics[width=4.5cm]{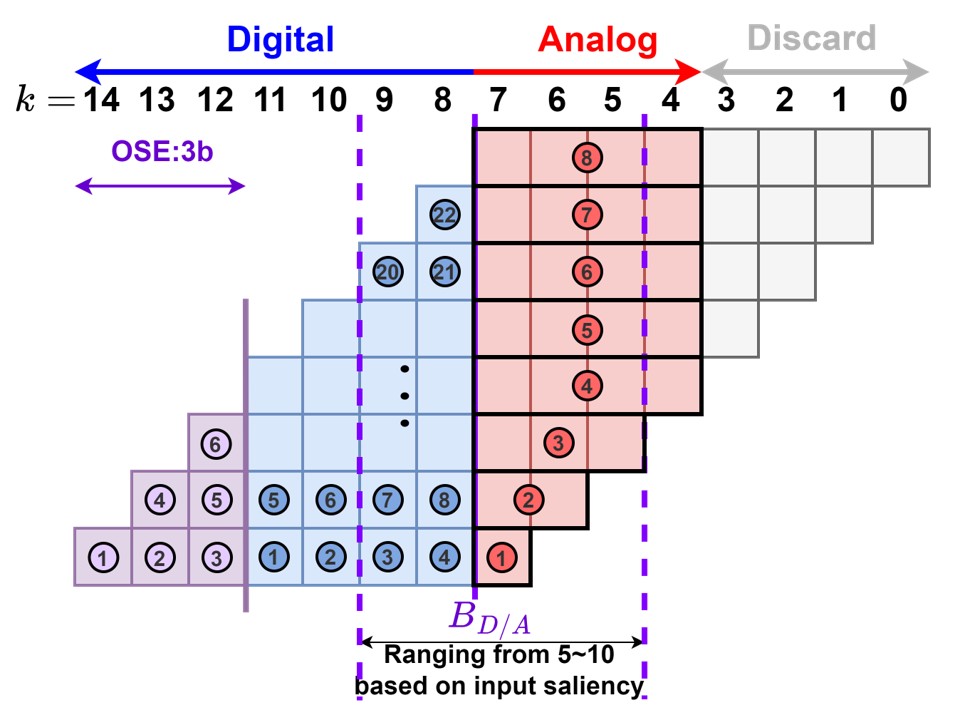}}%
\subfigure[]{%
\label{snr trade off}%
\includegraphics[width=4.5cm]{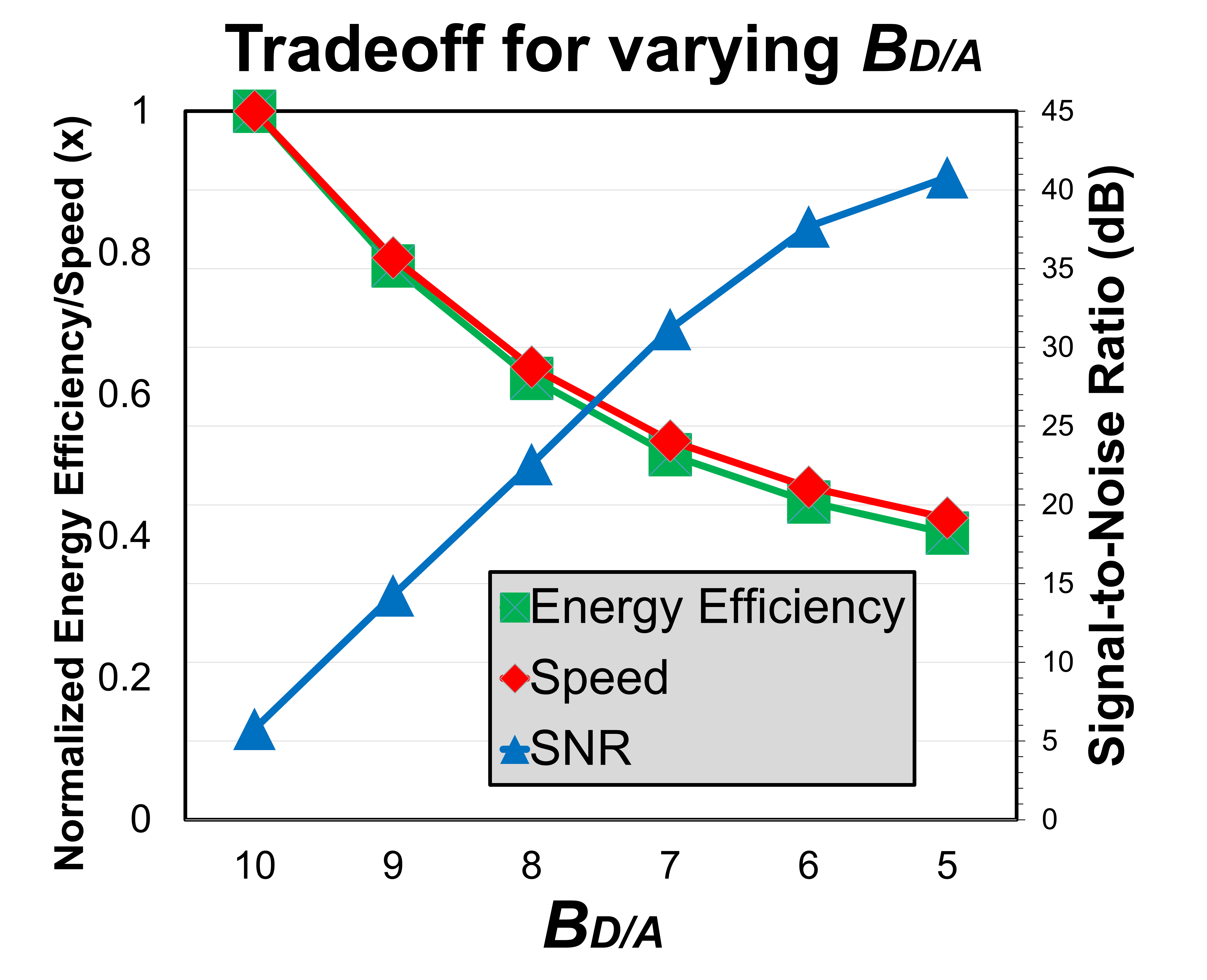}}%
\caption{(a) Workload allocation for DCIM and ACIM based on varying \bda. (b) Tradeoff between SNR, energy efficiency and execution speed under varying \bda{} for 8b $\times$ 8b MAC. The different Signal-to-Noise Ratio (SNR) and efficiency characteristics of the various \bda{} values meet the requirements for different types of input patterns, ensuring optimal performance across a range of scenarios. 
}
\vspace{-4mm}
\end{figure}

\subsection{On-the-fly Saliency Evaluator (OSE)}

The hardware implementation of the proposed On-the-fly Saliency Evaluator (OSE) is illustrated in Fig.\ref{OSE}. 
OSE plays a critical role in our framework --- it estimates the saliency and determines the suitable \bda{} of the multi-bit MAC operation from the high-order 1-bit MAC results. 
To achieve this, OSE first normalizes and quantizes the high-order 1-bit MAC result, denoted $DMAC_i$ in Fig.\ref{OSE}, of each channel. Then, these values are summed up and accumulated across cycles to obtain the saliency value $S$.
Based on the calculated saliency $S$, the OSE determines a suitable \bda{} from a candidate list $B = [B_0, B_1, ..., B_{b-1}]$. This is done by comparing $S$ with a set of thresholds $T = [T_0, T_1, T_{b-2}]$. This process is visualized in the histogram in Fig. \ref{OSE}. 

The thresholds $T$ are determined based on the algorithm illustrated in Fig. \ref{thres algo}. 
The algorithm is supplied with $B$ and a set of training loss constraints $L = [L_0, L_1, ..., L_{b-2}]$. Through an  iterative process that explores the threshold $T_i$ within the boundaries $B_i$ and $B_{i+1}$ to match the loss constraint $L_i$, a desired set of thresholds $T$ is obtained.
These thresholds are pre-trained, hence, they do not incur any additional overhead during the inference. Importantly, the loss constraints $L$ are specified by the user, allowing for customization of the desired tradeoff between accuracy and efficiency. This adaptability renders the framework compatible with a wide range of tasks. 


\subsection{Workload Allocation}

In this section, we propose a workload allocation to maximize the utilization of our hybrid CIM with digital and analog concurrent computation. 
Figure \ref{workload_allocate} illustrates the workload allocation for an 8b $\times$ 8b MAC. Based on the OSA precision configuration scheme, each 1-bit MAC is partitioned into three sections: \textit{digital}, \textit{analog}, or \textit{discard}. 
\textit{Digital mode} MACs are scheduled once per cycle using DCIM. Conversely, \textit{analog} 1-bit MACs with the same weight are calculated simultaneously in a bit-parallel manner using ACIM. Here, the variable bit-width is accommodated by the variable-precision DAC. Then, input activations are routed through GBLB/GBL, while weight access is facilitated by configuring DWL/AWL. 

Our scheme demonstrates its capability in providing a diverse selection of operating points with distinct precision and efficiency characteristics. As illustrated in Fig. \ref{snr trade off}, each value of \bda{} ranging from 5 to 10 represent a valuable operating point with distinct characteristics in terms of SNR and efficiency.
Compared to prior works that only utilize dual-precision for Saliency-Aware computing \cite{DRQ, PG, liu2020duet}, our methodology provides a significantly broader range of options to tailor to the varying saliency of each input.


Based on the allocation result, DCIM and ACIM compute the different assigned workloads concurrently, attaining an improved throughput.
One risk associated with the allocation scheme is the possibility of unbalanced workload between DCIM and ACIM, originated from the inherent throughput discrepancy and the variable \bda{} value.
To counteract this, DCIMs can be operated at a higher clock frequency, leveraging the fact that DAT has twice lower latency compared to ADC counterparts, since the SAR ADC requires 3 cycles to complete the conversion. 
\section{Simulation Results}

\newcommand{\ComparisonTable}{
\begin{table}[htbp]
\caption{Comparison with state-of-the-art SRAM CIM Macros}
\begin{center}
\begin{tabularx}{9cm}{
|>{\centering\arraybackslash}m{1.7cm}
|>{\centering\arraybackslash}X
|>{\centering\arraybackslash}X
|>{\centering\arraybackslash}X
|>{\centering\arraybackslash}m{1.5cm}
|}
\hline
& 
ICCAD '22\cite{lee2022lowcost} & 
ISSCC '21\cite{chih202116} & 
MCSoC '22\cite{chen2022charge} & 
\textbf{This Work}
\\
\hline
\textbf{Tech. (nm)} & 
28 & 
22 & 
22 & 
\textbf{65} \\
\hline
\textbf{CIM Type} & 
Analog & 
Digital & 
\makecell{
Fixed \\
Hybrid
} & 
\makecell{
\textbf{Dynamic} \\
\textbf{Hybrid}
} \\
\hline
\textbf{Input Prec.} & 
4b & 
1-8b & 
1b & 
\textbf{4/8b} \\
\hline
\textbf{Weight Prec.} & 
8b & 
4/8/12/16b & 
8b & 
\textbf{4/8b} \\
\hline
\textbf{Supply (V)} & 
0.6-1.2 & 
0.72 &  
0.8 &  
\textbf{0.6-1.2} \\
\hline
\textbf{Array Size} & 
256x64 &
256x256 & 
64x96 & 
\textbf{64x144} \\
\hline
\textbf{CIFAR100} & 
65.8\% &  
- & 
71.92\% &  
\textbf{67.4$\sim$72.1\%} \\ 

\textbf{Acc. (drop)$^{\mathrm{c}}$} & 
(0.5\%) &  
(0\%) & 
(4.17\%) &  
\textbf{(4.8$\sim$0.1\%)} \\ 
\hline
\textbf{ImageNet} & 
- &  
- & 
- &  
\textbf{65.2$\sim$70.8\%} \\ 

\textbf{Acc. (drop)} & 
(-) &  
(0\%) & 
(-) &  
\textbf{(6.3$\sim$0.8\%)} \\ 
\hline
\textbf{Energy Eff. (TOPS/W)}$^{\mathrm{a}}$ & 
5.7-22.9 & 
24.7 & 
6.98-11.0 & 
\multirow{2}{*}{
\makecell{
\textbf{5.33$\sim$5.79$^{\mathrm{d}}$} \\ @CIFAR100 \\
\textbf{3.83$\sim$4.66$^{\mathrm{d}}$} \\ @ImageNet 
}
}
\\

\cline{1-4}
\textbf{Norm. Energy Eff. (TOPS/W)}$^{\mathrm{b}}$
& 
1.06-4.25 & 
2.83 & 
0.80-1.26 & 
\\
\hline
\textbf{Saliency-Aware} & 
No & 
No & 
No & 
\textbf{Yes} \\
\hline

\multicolumn{5}{l}{$^{\mathrm{a}}$Normalized to 8b $\times$ 8b MACs (1 MAC is 2OPs).} \\
\multicolumn{5}{l}{$^{\mathrm{b}}$Normalized to 65nm process.}
\\ 
\multicolumn{5}{l}{$^{\mathrm{c}}$Compared to the baseline accuracy reported in each work.}
\\
\multicolumn{5}{l}{$^{\mathrm{d}}$Simulated under 0.6V supply voltage}
\end{tabularx}
\label{comparision_table}
\end{center}
\vspace{-4mm}
\end{table}

}

\begin{figure}[t]
\centerline{\includegraphics[width=8cm]{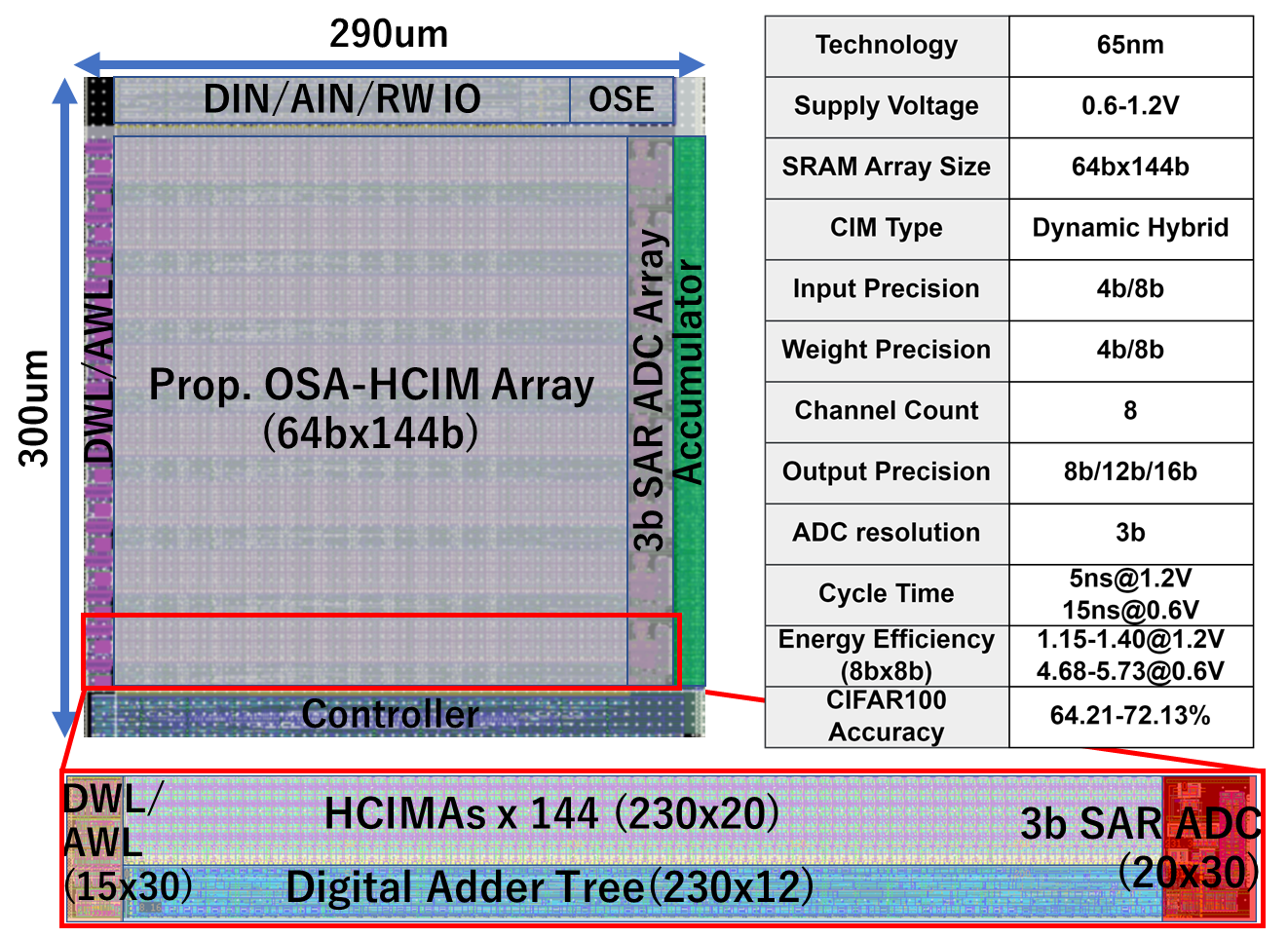}}
\caption{The layout of OSA-HCIM and its summary.}
\label{layout}
\vspace{-4mm}
\end{figure}

\begin{figure}[tbp]%
\centering
\includegraphics[width=7.5cm]{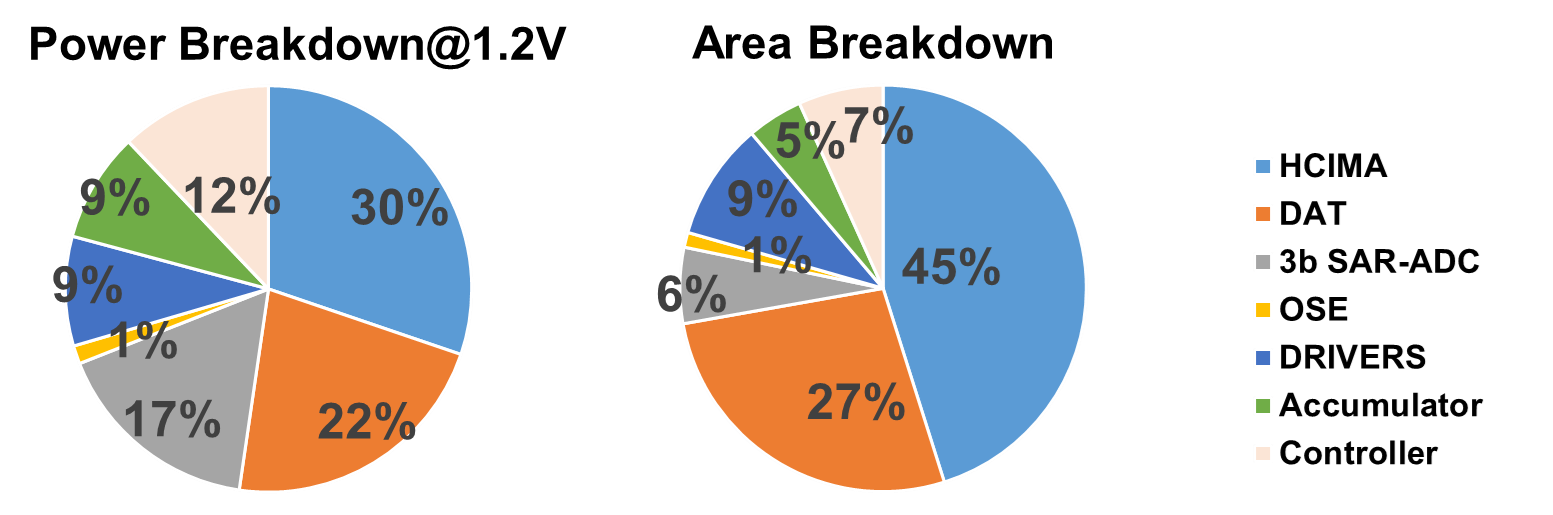}
\caption{
The power and area breakdowns of OSA-HCIM.
OSE incurs modest overhead thanks to the compressed input bandwidth and cost amortization across HMUs.
}
\label{breakdown}%
\vspace{-4mm}
\end{figure}

We implemented the 64b $\times$ 144b OSA-HCIM on a 65nm CMOS process. Figure \ref{layout} shows the layout and the summary of OSA-HCIM, and the power and area breakdowns of OSA-HCIM is shown in Fig. \ref{breakdown}. A significant advantage of our design is the OSE component--- despite its critical role in facilitating the OSA precision configuration scheme, it incurs only a minimal overhead of 1\% in power and 1\% in area, largely due to the compressed $DMAC$ bandwidth and cost amortization across 8 HMUs. Furthermore, unlike previous ACIM designs where the ADC was a primary power and area bottleneck, OSA-HCIM allows the utilization of low-precision 3-bit SAR ADC, which accounts for merely 17\% of the total power and 6\% of the total area. 

\begin{figure}[tbp]%
\centering
\subfigure[]{%
\label{saliency map}%
\includegraphics[width=4.2cm]{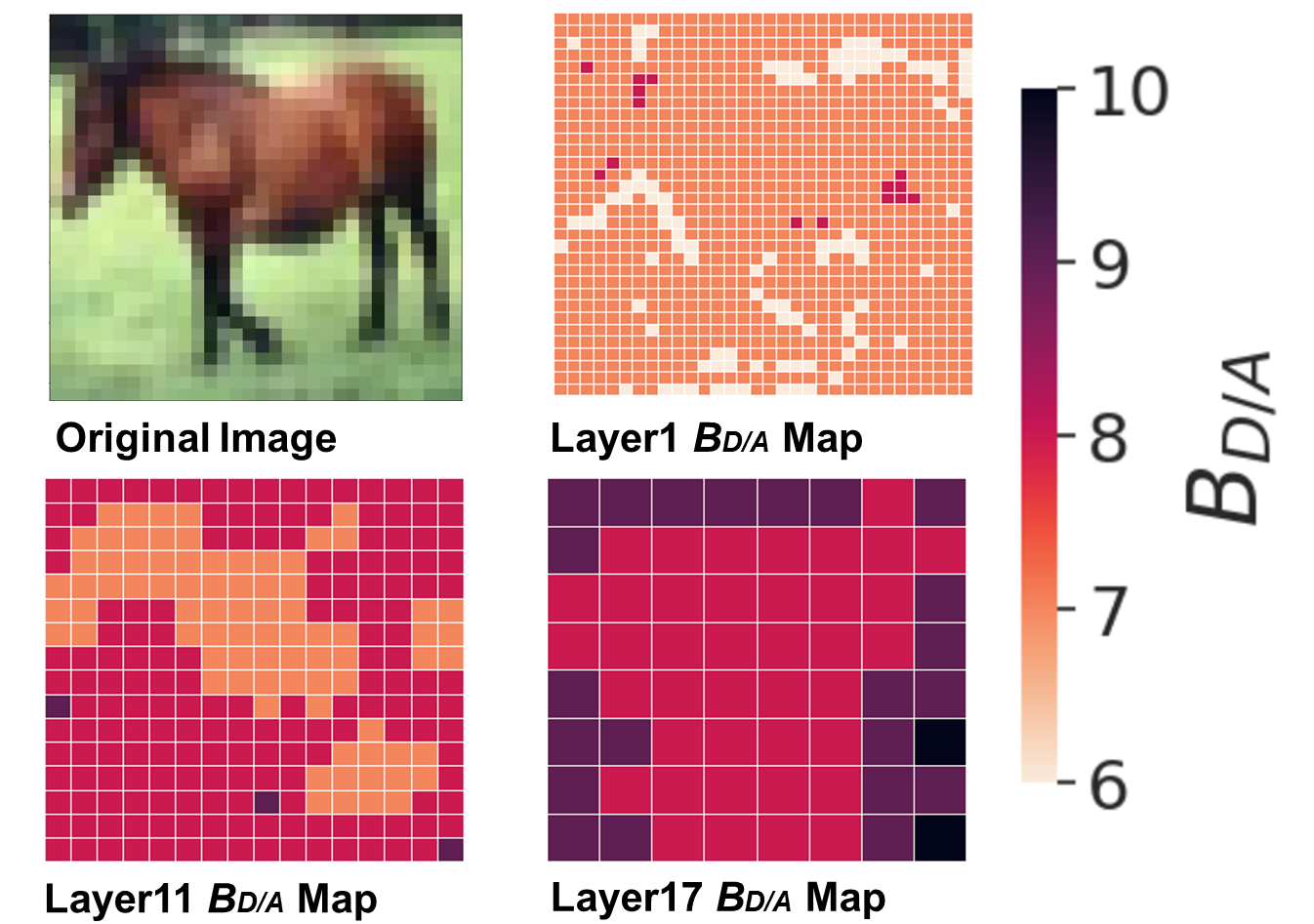}}%
\subfigure[]{%
\label{boundary ratio}%
\includegraphics[width=4.6cm]{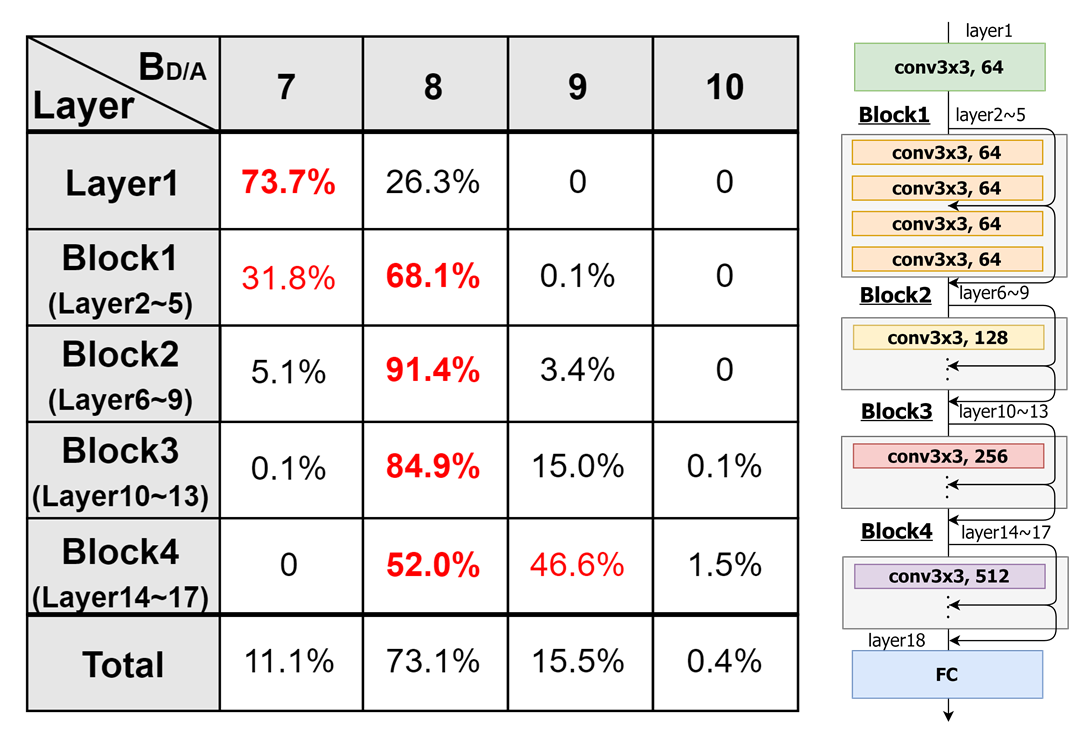}}%
\caption{(a) \bda{} maps of hidden layers for a horse image using ResNet20. OSE automatically maps high-precision computing settings to pixels pertaining the horse, and low-precision settings are assigned to background pixels. (b) The proportion of each \bda{} accross all CONV blocks of ResNet18 tested on CIFAR100 dataset. OSE effectively adapts to the optimum precision requirements across different layers.}
\vspace{-4mm}
\end{figure}

The \bda{} maps of various hidden layers processing a horse image using ResNet20 are shown in Fig. \ref{saliency map}. As observed in these layers, the majority of pixels pertaining to the horse are detected and assigned with relatively high-precision \bda{} settings, while background pixels are assigned to low-precision settings. These observations highlight the effectiveness of OSE in identifying salient pixels and adapting the fine-grained precision accordingly.
The adaptability of OSA-HCIM to saliency across layers is further evidenced in the analysis of \bda{} usage in ResNet18 for the CIFAR100 dataset, depicted in Fig. \ref{boundary ratio}. 
With the progression into deeper layers, there is a noticeable increase in the utilization of low-precision settings, effectively saving computational resources. The exceptional level of adaptability exhibited by OSA-HCIM, which exceeds that of previous works \cite{lee2021chargesharing8T}, achieves an optimal balance between accuracy and overall efficiency.

\begin{figure}[tbp]%
\centering
\includegraphics[width=8.5cm]{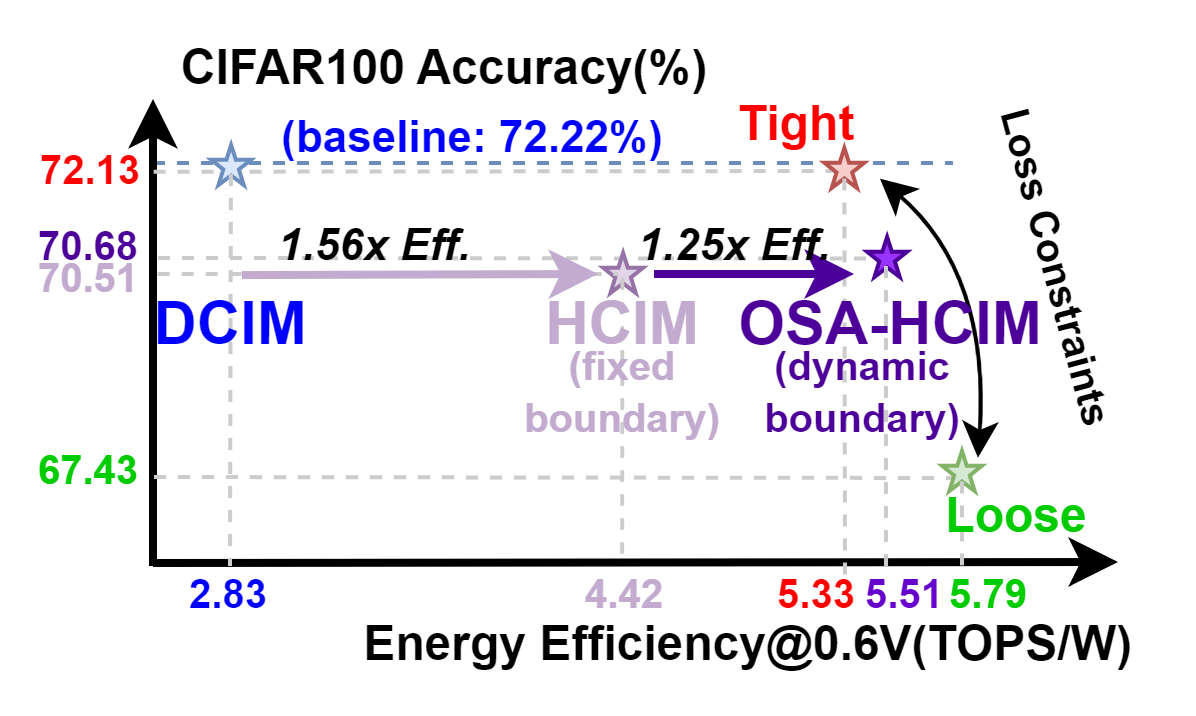}
\caption{By adjusting the loss constraints $L$, OSA-HCIM can adapt its operation to prioritize either higher accuracy or higher energy efficiency, depending on the specific requirements.
}
\label{tradeoff}%
\vspace{-4mm}
\end{figure}

Figure \ref{tradeoff} presents the correlation between CIFAR100 accuracy under ResNet18 and energy efficiency of DCIM, HCIM (without OSA), and OSA-HCIM. HCIM, which combines CIM of both analog and digital domains, achieves a 1.56x improvement in energy efficiency with $<2\%$ accuracy loss. OSA-HCIM, taking further advantage of input saliency, achieves an additional 1.25x efficiency boost, cumulating to a 1.95x total improvement compared to DCIM. In addition, OSA-HCIM offers a diverse set of operating points with unique accuracy-efficiency tradeoffs by adjusting the constraints $L$ in the threshold-finding algorithm in OSE, as discussed in Section V. It is capable of either obtaining $0.1\%$ accuracy drop or achieving high energy efficiency of 5.79 TOPS/W.
The flexibility afforded by the configurability of OSE empowers OSA-HCIM to address a variety of real-world tasks, each with its own precision requirements.
For instance, OSA-HCIM can achieve high accuracy even applied to more challenging datasets, such as ImageNet, since the algorithm in Fig. \ref{thres algo} autonomously explores thresholds that prioritize higher precision boundaries, with some tradeoff in efficiency.




Table \ref{comparision_table} presents the comparison with state-of-the-art designs. 
Notably, OSA-HCIM introduces the concept of Saliency-Aware Computing to CIM, enabling adaptive processing reflecting the input patterns. It is also the first CIM implementation to feature dynamic configuration of the digital-to-analog boundary, offering significant flexibility in terms of computing precision. Unlike conventional ACIMs, which often struggle to achieve high accuracy with large-scale datasets (e.g. CIFAR100, ImageNet), OSA-HCIM attains accuracy close to those of DCIM counterparts.
Moreover, OSA-HCIM delivers an energy efficiency of 5.33-5.79 TOPS/W, which is competitive against SRAM CIM designs when normalized to the 65nm process under CIFAR100 dataset.


\section{Conclusions}
Although CIM has been refined for general-purpose MAC operations, its efficiency suffers due to the lack of dynamic precision computing, which limits its adaptability to the saliency of different inputs. In response to this challenge, we introduced OSA-HCIM, a hybrid CIM architecture developed through software-hardware co-design.
The OSA precision configuration scheme enables the capability for online evaluation of input saliency, thus providing an array of precision settings tailored to different input patterns, enhancing computational accuracy and efficiency.
Furthermore, HCIMA, a CIM topology that enables simultaneous computation in the digital and analog domains, increases computational flexibility and boosts throughput.
The OSA-HCIM framework, bridging the OSA precision configuration scheme with HCIMA, exhibits remarkable versatility, effectively tackling tasks that demand varying precision requirements.
The proposed design, which is implemented using 65nm CMOS process, achieves a high energy efficiency of 5.33-5.79 TOPS/W while maintaining robust inference accuracy tested on the CIFAR100 dataset. 

{\footnotesize 
\section*{Acknowledgements}
This research was supported in part by the JST CREST JPMJCR21D2, JSPS Kakenhi 23H00467, Futaba Foundation, Asahi Glass Foundation, and the Telecommunications Advancement Foundation.}

\ComparisonTable

\bibliographystyle{IEEEtran}
\bibliography{mybib}

\vspace{12pt}
\end{document}